# Understanding Students' Acceptance, Trust, and Attitudes towards AI-generated Images for Educational Purposes

Aung Pyae

International School of Engineering, Faculty of Engineering, Chulalongkorn University, Bangkok, Thailand
aung.p@chula.ac.th

## Abstract

*Recent advancements in artificial intelligence (AI) have broadened the applicability of AI-generated images across various sectors, including the creative industry and design. However, their utilization in educational contexts, particularly among undergraduate students in computer science and software engineering, remains underexplored. This study adopts an exploratory approach, employing questionnaires and interviews, to assess students' acceptance, trust, and positive attitudes towards AI-generated images for educational tasks such as presentations, reports, and web design. The results reveal high acceptance, trust, and positive attitudes among students who value the ease of use and potential academic benefits. However, concerns regarding the lack of technical precision—where the AI fails to accurately produce images as specified by prompts—moderately impact their practical application in detail-oriented educational tasks. These findings suggest a need for developing comprehensive guidelines that address ethical considerations and intellectual property issues, while also setting quality standards for AI-generated images to enhance their educational use. Enhancing the capabilities of AI tools to meet precise user specifications could foster creativity and improve educational outcomes in technical disciplines.*

## 1. Introduction

Recent advancements in artificial intelligence (AI), such as deep learning and neural networks (Liang et al., 2022), have significantly impacted various sectors, including education. Notably, a research study by Fitria (2021) highlights how these technological improvements have boosted the efficiency of both teaching and learning, benefiting educators and students alike. Zhai et al. (2021) underscore the pivotal role of AI in education over the past decade, emphasizing the potential for collaborative endeavors between educators and AI developers. Furthermore, the introduction of AI tools such as ChatGPT, Gemini, and Copilot has not only enhanced educational practices but also sparked ongoing ethical debates around issues such as plagiarism, highlighting the importance of balancing AI's benefits with ethical considerations (Adeshola & Adepoju, 2023).

In recent years, AI technologies like OpenAI's DALL-E have revolutionized image creation through advanced machine learning models. These models include Generative Adversarial Networks (GANs), which use a competitive process between two networks to generate realistic images, and Variational Autoencoders (VAEs), which encode and decode data to produce new variations of images. These systems generate visually compelling content that closely mimics their training data, demonstrating a significant stride in AI capabilities, rooted in the foundational research by Goodfellow et al. (2014) and Ramesh et al. (2021). While AI-generated images promise diverse applications, they also pose ethical challenges, particularly concerning authenticity and misuse, which must be addressed for responsible usage, as highlighted by Zhu et al. (2021).

Recent studies on AI-generated images have broadened their applications, with researchers increasingly interested in the quality, authenticity, and overall usefulness of these images in various fields. Chen et al. (2024) introduced *'ArtScore,'* a metric for assessing the artistic quality of AI-generated images that closely aligns with human judgments, marking a significant advancement in the evaluation of digital artistry. Similarly, Zhu et al. (2023) developed the *'GenImage'* dataset, designed to combat disinformation by effectively distinguishing between real and AI-generated images, enhancing the integrity of digital media. Further contributing to this field, Epstein et al. (2023) successfully differentiated between traditional and AI-generated images, including those with mixed components, providing insights into the complexities of digital image verification. Additionally, Baraheem and Nguyen (2023) utilized a CNN framework to refine image analysis techniques, achieving high accuracy in identifying synthetic images and further demonstrating the capabilities of advanced machine learning in the realm of image authentication.

Together, these studies reflect the dynamic evolution and growing impact of AI in the field of image generation and analysis. They highlight not only the technological advancements but also address the critical need for tools that can ensure the ethical use of AI-generated visuals in various applications. This body of work sets the stage for ongoing research that could transform how digital images are created, analyzed, and utilized across multiple disciplines.

Despite extensive research on the technological and ethical aspects of AI-generated images, their specific applications in educational settings, particularly from the student perspective, remain underexplored. Previous studies have primarily focused on AI's technical development from the viewpoints of developers, missing direct insights from student interactions in educational settings. This gap is significant as understanding student engagement with AI technologies could provide crucial insights into their practical benefits, challenges, and ethical implications within educational environments. This study investigates how AI-generated images can enhance learning experiences in activities like report writing, presentations, and web design, promoting experiential, collaborative, and visual learning. It examines university students' usage, perceptions, acceptance, trust, and attitudes towards these AI tools, aiming to understand the broader implications of integrating AI in image generation for enhancing educational practices. This research advocates for a more effective use of AI technologies in learning environments, aiming to improve educational outcomes and foster a deeper integration of these tools in academic settings.

## 2. Related Studies

The article by Rubman (2024) in the MIT Sloan Review classifies AI-generated images in educational contexts into three distinct categories: instructive, decorative, and distracting. It emphasizes that these images can significantly enhance learning when strategically chosen to align with educational goals and foster inclusivity. Further, the Apple Education Community (2024) underscores the transformative potential of AI text-to-image technologies in children's literature. It highlights their capacity to deliver consistent, high-quality visuals that enrich storytelling and learning experiences, thus setting a new standard for digital educational resources.

Moreover, a study by Thorburn (2024) from the Hong Kong TESOL explores the practical applications of AI image generators in education. It illustrates how educators can swiftly create customized teaching aids, such as flashcards and visual aids, that are directly tied to lesson themes and localized contexts. This approach not only facilitates engaging learning activities like "spot-the-difference" games but also circumvents the usual copyright restrictions associated with traditional internet images.

Building on these insights, Aktay (2022) investigates the efficacy of these tools through a qualitative assessment using OpenAI's DALL-E. The findings suggest a robust capability of AI to visualize both abstract and concrete concepts, opening up promising avenues for educational applications. Complementarily, Warner (2024) introduces innovative methods for embedding AI-generated images into language instruction, proposing five strategic exercises that bolster visual engagement and enhance language skills. In a similar vein, Goldman (2024) examines the use of AI-generated images in preparing teachers for special education. The study proposes that such tools could revolutionize teaching methodologies by providing tailor-made visuals that address specific educational challenges in special education settings. Expanding this discourse to a broader spectrum, Ringvold et al. (2023) analyze the implications of AI text-to-image generators in art and design teacher education, drawing parallels with the disruptive historical impact of the photographic camera. They discuss how AI redefines traditional visualization skills and poses both opportunities and challenges in educational frameworks.

The World Economic Forum (2024) adds to the discussion by emphasizing the transformative potential of generative AI across educational sectors, advocating for its thoughtful integration to redefine conventional teaching and learning paradigms and enhance personalized educational experiences. Concurrently, Reed et al. (2023) explore the perceptions and concerns of undergraduate nursing students regarding the use of Midjourney's generative AI in their curriculum, highlighting its therapeutic and technological advantages as a pedagogical tool.

Despite these advancements, the literature identifies areas of concern, such as the study by Temsah et al. (2024), which points out significant inaccuracies in AI-generated images used in medical education for depicting congenital heart diseases. This underlines the critical need for caution in their educational application. These collective studies illuminate the varied potentials and challenges of AI in diverse educational fields. Yet, they also reveal significant gaps in understanding students' acceptance, trust, and attitudes toward AI-generated images, as well as their actual usefulness in educational settings. This underscores an urgent need for more comprehensive studies to assess the instructional effectiveness of these images and ensure their utilization truly enhances learning without introducing distractions or inaccuracies. Given these identified gaps, this study sets forth three objectives: 1) to assess students' acceptance of AI-generated images for educational purposes, 2) to explore their trust and attitudes towards these images, and 3) to evaluate the usefulness of AI-generated images in education.

## 3. Method

The usability testing was conducted in a university lab setting with 15 undergraduate students from the Software Engineering and Information Technology programs, all aged between 18 and 25. Recruitment was facilitated through the university's social media student groups, with participants meeting specific inclusion criteria: current enrollment in the institution and digital literacy. Comprehensive demographic data, including age, gender, academic year, and previous experience with AI tools, were collected to enhance the analysis. A pilot test involving two students was initially conducted to refine the study design and procedures. Feedback from this pilot led to minor revisions in the clarity of the prompts and interview questions, aiming to enhance the overall effectiveness of the study. The primary aim of this testing was to evaluate students' acceptance, trust, and attitude towards AI-generated images for educational tasks such as report preparation and creative design.

During the in-study phase, participants used ChatGPT 4.0, accessed through the latest version of the Chrome browser on standardized laptops with high-speed internet to ensure consistency across sessions. They generated images based on pre-defined prompts that were developed after extensive discussions with experts in computer science, psychology, and education. These prompts were categorized under ten themes, including technology, arts, nature, and health, designed to evaluate the AI's versatility in creating educational images comprehensively (see Table 1). Participants engaged with prompts from ten different themes, spending a total of about 30 minutes critically evaluating the AI-generated images based on criteria such as relevance, creativity, and accuracy, and documenting their qualitative assessments. To minimize potential order effects and ensure a robust assessment of each theme, the order of prompts was randomized for each participant in the study.

In the post-study phase, participants completed questionnaires and participated in interviews to assess the psychological and behavioral aspects of adopting new technologies. They used the Technology Acceptance Model (TAM), adapted from Aburbeian et al. (2022), which evaluates self-efficacy, social norms, and perceived usefulness, among others. Davis (1989) notes that TAM is apt for our study on AI-generated images for students, as it assesses user acceptance and utilization. The Trust Scale, adapted from Merritt (2011), measured trust in AI-generated imagery, vital for incorporating educational technology. Additionally, the Technology Attitude Scale (TAS), from Rosen et al. (2013), assessed attitudes towards these images, focusing on confidence and perceived benefits. Both TAM and the Trust Scale used a 5-point Likert scale, while TAS employed a 6-point scale. Open-ended interviews offered further qualitative insights into students' perceptions of AI-generated images in education. Data analysis included descriptive and thematic methods, providing a comprehensive understanding of student acceptance. The entire study required about 45 minutes per participant. Ethical considerations were meticulously adhered to throughout the study, with protocols ensuring participant consent, confidentiality, and voluntary participation. Table 2 shows the detailed usability test design and procedures.

**Table 1. Prompts used in the usability test.**

| Themes | Prompts |
|---|---|
| Technology | Create an image of a futuristic cityscape with autonomous vehicles, digital billboards, and smart buildings powered by renewable energy, depicting advanced urban living. |
| Arts and Culture: | Create an image of a traditional Thai dance, featuring dancers in elaborate costumes and masks at a historic temple, capturing Thai cultural heritage. |
| Nature and Environment: | Create an image of a lush rainforest with diverse flora and fauna, a waterfall, and a serene river, emphasizing ecosystem preservation. |
| People | Create an image of Asian friends from various countries sharing a meal in a city park, depicting cultural exchange and friendship. |
| Travel | Create an image of a traveler overlooking the majestic Himalayas from a viewpoint, embodying the spirit of adventure and exploration. |
| Health and Wellness: | Create an image of a serene outdoor yoga session in a tranquil garden with people of all ages practicing mindfulness, emphasizing mental and physical health. |

| Religion | Create an image of a peaceful, ornate Buddhist temple in Thailand during meditation, with monks and visitors in a tranquil setting, reflecting its spiritual and architectural beauty. |
|---|---|
| Universe | Create an image of an astronaut observing a distant galaxy from a spacecraft window, capturing the magnificence and mystery of space. |
| Animals | Create an image of a panda family in a bamboo forest in China, highlighting their playful interactions and the serene beauty of their habitat. |
| Games and Rewards: | Create an image of people enjoying a modern gaming experience with futuristic VR equipment and an immersive game world, highlighting the excitement of digital entertainment. |

**Table 2. Usability testing design.**

| Study | Task | Duration |
|---|---|---|
| Pre-study | • Introduction to the study<br>• Informed consent<br>• Pre-study questions | 10 mins |
| In-study | • AI image generation using the pre-defined 10 prompts | 30 mins |
| Post-study | • Questionnaires: TAM, Trust Scale, Technology Attitudes Scale<br>• Interview | 30 mins |

## 4. Results and findings

### 4.1. Pre-study questions

The study involved 15 students aged 18-25, with a gender distribution of 53.33% female (8) and 46.67% male (7). The use of AI-generated images for academic purposes varied among participants: 33.3% used them moderately, 26.7% rarely, 20% infrequently, and 20% frequently. No participants reported very frequent use. The majority (66.7%) employed these images in presentations, 46.7% in assignments or reports, and fewer in work-related tasks and social media. Notably, none utilized them for thesis projects. Canva was the most popular tool for creating images, used by 66.7% of participants, followed by Google Images (60%) and Photoshop (13.3%). Other platforms like DALL-E and Freepik saw minimal use. Regarding perceived usefulness, nearly half of the students found the images moderately useful, one-third highly useful, and the remainder perceived limited utility. These results indicate that while AI-generated images are helpful, they are not deemed essential for academic success. Figure 1 shows the selected images generated by AI during the usability sessions.

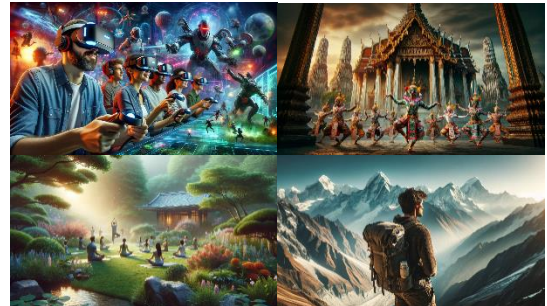

**Figure 1. AI-generated images.**

### 4.2. Technology acceptance

This study's examination of technology acceptance for AI-generated images revealed multiple dimensions of user interaction, emphasizing high self-efficacy with participants reporting strong capabilities in using AI tools for image generation (TA1: $M = 4.1$, $SD = 0.6$; TA3: $M = 4.5$, $SD = 0.5$), and a minimal perceived need for specialist support (TA2: $M = 1.6$, $SD = 0.8$). Social norms exerted a moderate influence on usage (TA4: $M = 2.4$, $SD = 0.9$; TA5: $M = 3.5$, $SD = 1.18$), while curiosity and pleasure were highly rated (TA6: $M = 3.3$, $SD = 0.8$; TA7, TA8: $M = 3.9$, $SD = 1.0$; TA9: $M = 4.2$, $SD = 0.5$), indicating significant engagement and enjoyment among users. Notable concerns about cost (TA10: $M = 3.7$, $SD = 1.0$) suggest the need for effective pricing strategies. The high marks for perceived usefulness and ease of use (TA11: $M = 4.1$, $SD = 0.8$; TA12: $M = 3.3$, $SD = 1.2$; TA13: $M = 4.5$, $SD = 0.5$), coupled with strong behavioral intentions (TA14: $M = 4$, $SD = 0.8$) and positive attitudes (TA15: $M = 3.9$, $SD = 0.5$), support a promising outlook for the integration of AI-generated images in educational settings. This comprehensive data suggests that while users are adept and engaged with AI tools and images, addressing cost concerns could further enhance adoption rates.

## 4.3. Trust scale

The results indicate varied trust levels in AI-generated images across six dimensions. Competence in creating AI images was rated at a mean of 3.6 with a standard deviation of 0.6 (M=3.6, SD=0.6), suggesting a general perception of proficiency in the AI creation process. Overall trust in these images showed a foundational level of confidence (M=3.3, SD=0.7). Confidence in their performance was slightly higher (M=3.4, SD=0.6), with dependability similarly rated (M=3.3, SD=0.7). Behavioral consistency and reliability in performance both received higher trust ratings, each scoring M=3.7 with standard deviations of 0.8 and 0.6, respectively. These results reflect a moderate to high trust in AI-generated images and underscore areas where further improvements could enhance user adoption and engagement.

## 4.4. Technology attitude scale

The Technology Attitude Scale (TAS) was utilized to assess participants' attitudes toward AI-generated images, focusing on confidence, perceived benefits, and integration into academic and future professional settings. High mean scores indicated positive attitudes and confidence: importance of knowing how to use AI images (TAS1: M=4.1, SD=1.1), enjoyment of usage (TAS2: M=4.6, SD=0.6), confidence in learning about AI images (TAS3: M=4.5, SD=0.8), and recognition of learning value (TAS4: M=4.7, SD=0.8). Participants showed a strong inclination to apply this knowledge academically (TAS6: M=4.7, SD=0.8) and recognized its importance for future careers (TAS8: M=4.6, SD=0.9). The facilitative role of AI images in enhancing learning was acknowledged (TAS10: M=4.4, SD=0.9), although certainty about the direct impact on academic performance was less pronounced (TAS11: M=3.9, SD=1.0; TAS13: M=3.6, SD=1.12). Negative sentiments were low, with minimal feelings of inadequacy (TAS7: M=1.7, SD=0.7) or difficulty in using AI images (TAS12: M=1.5, SD=0.6), and discomfort was the lowest score (TAS14: M=1.3, SD=0.4). Overall, the TAS results depict a robustly positive attitude toward AI-generated images, highlighting significant confidence and perceived utility, with minimal associated anxiety or discomfort.

## 4.5. Interview analysis

Thematic analysis of student feedback from structured interviews on AI-generated images in academic settings revealed key themes as follows:

**4.5.1. Perceived quality of AI-generated images**. The theme "*Perceived Quality of AI-Generated Images*" prominently emerged from the participants' feedback, focusing on the aesthetic and practical qualities of AI-generated images for academic use. The majority appreciated the image quality, with one participant noting, "*95% of the generated images are in good quality*," indicating high satisfaction. Another participant emphasized the convenience, stating the images "*can be readily used*" without further editing, highlighting their immediate usability and standards compliance. The participants praised the images as "*very beautiful*" and "*colourful*," underscoring their visual appeal which is vital in academic use. Despite positive feedback on aesthetics, concerns about the realism of AI-generated images highlight a gap between their visual appeal and accurate real-world representation, indicating a need for improvement in educational and professional contexts.

**4.5.2. Realism and accuracy concern**. The theme "*Realism and Accuracy Concerns*" emerged as a critical aspect of feedback on AI-generated images, particularly in their portrayal of human figures. Participants noted a significant gap between the images' aesthetic appeal and their realism, with one stating, "*Human images are very much like AI-generated content, not realistic*," highlighting the artificial appearance, especially in body parts other than faces, likened to paintings. Additional concerns included textual inaccuracies, such as misspelled words or illegible text, as one participant noted, "*The main issue is that it cannot produce correct and accurate texts in the image itself*," and unclear backgrounds and contextual elements. This feedback highlights the need for substantial improvements in the realism and accuracy of AI-generated images, crucial for applications demanding precise human representation and textual clarity, particularly in educational and professional settings.

**4.5.3. Utility for academic purposes**. The theme "*Utility for Academic Purposes*" reflects students' appreciation for the educational benefits of AI-generated images. Many participants recognized their practicality, particularly in creative and design tasks. For example, one participant mentioned the utility of these tools for students who study design (e.g., graphics and websites) to create draft drawings and generate ideas. Additionally, the effectiveness of AI-generated images in enhancing presentations and reports was noted, with comments on their high quality and immediate usability for university assignments. However, while the participants valued these images for educational uses, they also expressed concerns about their limitations in more professional or technically demanding contexts, noting a gap in technical precision. This feedback underscores the need for further refinement of AI-generated images to meet higher standards of accuracy and professionalism.

**4.5.4. Technical limitations and specificity**. The theme "*Technical Limitations and Specificity*" highlights the participants' frustration with the technical shortcomings of AI-generated images, particularly in tasks requiring high precision. Students noted that AI images are unsuitable for technical purposes, such as producing accurate technical diagrams, indicating a significant limitation in detail and specificity. Feedback also pointed to the AI's inability to consistently interpret and execute detailed user prompts accurately, with students reporting that the outputs often do not match the requested specifications. These issues underscore the need for improvements in AI's understanding and execution capabilities to enhance its utility in specialized academic and professional applications.

**4.5.5. Future usage intentions**. The theme "*Future Usage Intentions*" reflects students' optimistic outlook on the role of AI-generated images in future academic projects, despite current limitations. Students are hopeful about the technology's potential improvements in realism and technical precision. For example, one participant expressed a conditional willingness to use AI images in the future as the technology improves, while another planned to use them for educational purposes, contingent on advancements in quality. Additionally, one participant mentioned the potential usefulness of AI images for colorful cover designs, emphasizing their growing utility in diverse educational contexts. This forward-looking sentiment underscores the dynamic nature of AI technology and the expectation that ongoing enhancements will address existing deficiencies, particularly in realism and the ability to meet specific prompts. This optimism highlights the potential of AI-generated images in academic settings and underscores the need for continuous innovation in AI technologies.

**4.5.6. Suggestions for improvement**. The theme "*Suggestions for Improvement*" reflects students' constructive feedback on enhancing AI-generated images, particularly in achieving more natural and realistic human representations. The Participants emphasized the need for AI to produce images that appear less artificial, with comments like, "*The human images should be more natural; as currently they are too AI or fake*." Additionally, there is a demand for AI to more accurately interpret and execute user prompts, ensuring that outputs align closely with specific requirements. The participants also suggested increasing the variety and customization in the image generation process, including options to produce different versions of an image and to improve aspects such as color saturation. These improvements are aimed at offering more flexibility and precision, enhancing the overall utility and appeal of AI-generated images in academic and various other contexts.

**4.5.7. Creative freedom and exploration**. The theme "*Creative Freedom and Exploration*" highlights how AI-generated images serve as enablers of creativity and innovation, extending beyond mere functional utilities. Students emphasized the potential of these tools to enhance creativity, with one noting, "*Using AI-image tools, you can be more creative*," suggesting that these tools encourage experimentation and originality. Additionally, AI-generated images are seen as sources of inspiration, particularly useful in fields like design and architecture for sparking new ideas. This perspective underscores the significant role of AI in fostering creative expression and ideation, opening new avenues for innovation in academic and professional projects, especially in disciplines that prioritize novel visual concepts.

**4.5.8. Flexibility and specific applications**. The theme "*Flexibility and Specific Applications*" reflects the participants' recognition of the specialized utility of AI-generated images for particular project types, highlighting the adaptability of AI tools to meet diverse academic and professional needs. For instance, one participant noted the advantage of AI-generated images for network design projects, appreciating its capability to provide specific visual representations. Additionally, the versatility of AI images in web development was emphasized, with participants pointing out its significant role in enhancing web design through visually appealing and appropriate graphics. The utility of AI tools in generating icons also demonstrates their value in creating crucial visual elements for various designs and digital contexts. This theme underscores the perception of AI-generated images as a versatile tool that can be tailored to a wide range of specific requirements, showcasing the strength of AI in providing customized solutions that improve the quality and relevance of visual content across different fields.

**4.5.9. Other Findings**. The thematic analysis of the participants' feedback on AI-generated images highlighted considerations beyond technical aspects, such as cost, accessibility, customization, and legal concerns. Affordability and easy access are key factors influencing the use of AI tools. Students expressed a desire for greater customization and control, advocating for more interactive features to personalize the image creation process, such as photo edit settings and the ability to produce different image versions. Additionally, AI-generated images are viewed as advantageous over traditional methods, particularly for avoiding copyright issues, and offering convenience and legal safety. This analysis reveals that students' considerations for using AI images are complex, involving economic, practical, and legal factors.

In summary, thematic analysis of student feedback on AI-generated images in academic settings highlighted their appreciation for aesthetic qualities but identified gaps in realism and technical precision. Key themes included perceived quality, utility for academic purposes, and future usage intentions, with students expressing optimism for improvements and a strong interest in using AI images for creative and specific applications. This underscores the need for continuous innovation in AI image technologies.

## 5. Discussion

In this study, the findings reveal complex yet interesting responses from the participants, characterized by robust engagement and notable recognition, alongside identified challenges and opportunities for future enhancements. The findings show that students demonstrated high self-efficacy, using AI tools independently, which is crucial for their broader acceptance in both educational and professional environments. While social norms moderately influence adoption, intrinsic curiosity strongly drives usage, indicating significant opportunities for innovation and increased adoption.

The exploratory study's results indicate that the students find AI-generated images useful and enjoyable, enhancing their user experience. Economic factors (e.g., cost) emphasize the need for making these tools more accessible to promote wider educational use. Trust in AI-generated images is positive, with users confident in its performance and reliability, suggesting a promising future. Furthermore, students' positive attitudes and minimal anxiety about using AI-generated images indicate a trajectory for growth by enhancing trust and ease of use. Overall, curiosity, utility, and user satisfaction create a strong foundation for the ongoing development and acceptance of AI-generated image technologies in academia, which is in line with the existing literature (Fitria, 2021).

However, despite these favorable outcomes, student feedback also reveals important concerns that need addressing to fully capitalize on the potential of AI-generated images in educational settings. In the analysis of student feedback on AI-generated images in educational settings, there emerges a nuanced balance between their aesthetic appeal and practical utility, juxtaposed with concerns about realism and technical accuracy. Students value the images' visual appeal for enhancing educational engagement but highlight realism gaps, especially in human depictions, underscoring a need for more accurate portrayals. Despite aesthetic benefits, their practical utility is limited in precision-demanding contexts. Students remain optimistic about future enhancements in AI imagery that could expand its educational applicability. They recognize the necessity for AI tools to be more adaptable, user-friendly, and economically accessible. As AI technology evolves, it promises to increasingly enrich educational experiences, provided ongoing innovations address current challenges in realism, accuracy, and specificity (Temsah et al., 2024).

In addressing Objective #1, our study aimed to assess students' acceptance of AI-generated images for educational purposes, focusing particularly on user perceptions and self-efficacy concerning this technology. The findings revealed that users feel empowered and comfortable using AI-generated images, driven by a strong intrinsic motivation to integrate these tools into educational contexts. This enthusiasm underscores a potential for sustainable growth in the use of AI-generated images within educational settings, propelled by a genuine interest in leveraging the technology for innovative purposes.

Moving to Objective #2, we assessed trust and attitudes towards AI-generated images in education, noting moderate to high trust based on the technology's perceived reliability and performance. Positive user experiences significantly contribute to this trust, emphasizing the need for continual improvements to sustain its success. However, concerns about the realism and accuracy of the images, particularly in human depictions, call for further technological refinement.

Finally, Objective #3 evaluated the utility of AI-generated visuals in education. While effective for tasks like presentations, these images struggle in precision-heavy contexts, underscoring the need to balance aesthetics with accuracy—essential in educational settings. Optimism for future improvements is tempered by current realism and accuracy issues. The study stresses ongoing AI advancements to improve educational applications and suggests that broader, more diverse research is needed to confirm these findings and unlock the full potential of AI-generated images.

## 6. Conclusion

In conclusion, this study highlights the strong engagement and self-efficacy of students with AI-generated images in educational contexts, showing high acceptance of the technology. While positive attitudes and trust are prevalent, concerns over realism, accuracy, and technical limitations suggest areas for enhancement. The potential of AI-generated images to improve educational materials depends on advancements in realism, precision, and customization. Future developments in AI and demand for adaptable, affordable tools indicate significant opportunities for broader integration and innovation in education.